\documentclass[conference]{IEEEtran}
\IEEEoverridecommandlockouts

\usepackage[backend=biber,style=ieee]{biblatex}
\def\BibTeX{{\rm B\kern-.05em{\sc i\kern-.025em b}\kern-.08em
    T\kern-.1667em\lower.7ex\hbox{E}\kern-.125emX}}

\usepackage{amsmath,amssymb,amsfonts}
\usepackage{algorithmic}
\usepackage{graphicx}
\usepackage{textcomp}
\usepackage{xcolor}
\usepackage{braket}
\usepackage{qcircuit}
\usepackage{subcaption}

\begin{document}

\title{Variational Quantum Models for Knowledge Graph Embeddings on NISQ Devices}

\makeatletter
\newcommand{\linebreakand}{%
  \end{@IEEEauthorhalign}
  \hfill\mbox{}\par
  \mbox{}\hfill\begin{@IEEEauthorhalign}
}
\makeatother

\author{\IEEEauthorblockN{Guido Bellomo}
\IEEEauthorblockA{\textit{Instituto de Ciencias de la Computación} \\
\textit{CONICET-Universidad de Buenos Aires}\\
Buenos Aires, Argentina \\
gbellomo@icc.fcen.uba.ar}
\and
\IEEEauthorblockN{Martín Santesteban}
\IEEEauthorblockA{\textit{Dept. de Computación} \\
\textit{Universidad de Buenos Aires}\\
Buenos Aires, Argentina \\
martin.p.santesteban@gmail.com}
\and
\IEEEauthorblockN{Patricio Bruno}
\IEEEauthorblockA{\textit{Dept. de Computación} \\
\textit{Universidad de Buenos Aires}\\
Buenos Aires, Argentina}
\linebreakand
\IEEEauthorblockN{Santiago Cifuentes}
\IEEEauthorblockA{\textit{Dept. de Computación} \\
\textit{Universidad de Buenos Aires}\\
Buenos Aires, Argentina}
\and
\IEEEauthorblockN{Gustavo Martin Bosyk}
\IEEEauthorblockA{\textit{Instituto de Ciencias de la Computación} \\
\textit{CONICET-Universidad de Buenos Aires}\\
Buenos Aires, Argentina \\
gbosyk@icc.fcen.uba.ar}
}

\maketitle

\begin{abstract}
Variational Quantum Algorithms (VQAs) combine quantum circuits with classical optimization to tackle problems that may benefit from the capabilities of near-term quantum hardware. In knowledge graph embedding, recent proposals based on this approach follow a similar overall architecture but differ in the way they compute the score function and in the number of qubits they require. One design uses $n+1$ qubits and obtains the score through a switch test on an ancillary qubit, while another employs $2n+1$ qubits and applies a swap test between two registers. In both cases, entities and relations are represented in a Hilbert space of dimension $d = 2^n$, with comparable computational cost and the same mean squared error loss. This work introduces a unified framework that captures the two schemes and makes it possible to explore new variants. Within this setting, we propose an alternative that keeps the intuitive meaning of the score function while dispensing with ancillary qubits and entangled measurements. The result is a model better suited to current NISQ devices, reducing hardware demands without sacrificing interpretability.
\end{abstract}


\section{Introduction}
Knowledge graphs (KGs)~\cite{Hogan2022} have emerged as key structures in contexts where the relationships between data points are as important as the data itself. Applications range from social networks~\cite{fan2012graph} to the Semantic Web~\cite{arenas2011querying} and data provenance~\cite{anand2010techniques}. Mathematically, a KG is a directed multigraph $KG=(V, R, E)$, where $V$ is the set of entities, $R$ the set of relation types, and $E \subseteq V \times R \times V$ the set of triples $(h,r,t)$ expressing that head entity $h$ is related to tail entity $t$ via relation $r$.

Common KG tasks such as data completion~\cite{rossi2021knowledge}, clustering~\cite{saeedi2018using}, and link prediction~\cite{kumar2020link} are often addressed through embedding-based methods~\cite{ge2024knowledge, dai2020survey}. These approaches map each entity $e \in V$ and each relation $r \in R$ to vectors in a low-dimensional space such that similar embeddings reflect shared structural semantics. A score function $\delta_{hrt}: V \times R \times V \to \mathbb{R}$ estimates the plausibility of each triple $(h,r,t)$, and model parameters $\Theta$ are optimized on a training set $\mathcal{D} = \{(h,r,t), y_{hrt}\}$—with $y_{hrt}$ indicating whether the triple belongs to the KG—by minimizing a loss function $\mathcal{L}(\delta_{hrt}, y_{hrt}; \Theta)$. Classical methods such as RESCAL~\cite{nickel2011three}, TransE~\cite{asmara2023review}, and RotatE~\cite{sun2018rotate} follow this paradigm.

\subsection{Quantum embeddings for KGs}

Recently, quantum methods have started to explore how Hilbert space can be used for learning KG embeddings. VQAs are one such approach, combining parameterized quantum circuits with classical optimization. Thanks to their shallow circuits, VQAs work well on current noisy quantum devices. When applied to KGs, they represent entities and relations as quantum states or unitaries and use measurements to calculate scores. This provides an alternative way to represent relational data, which may help identify patterns that classical models struggle with.

\begin{figure}[!t]
  \centering
  \begin{subfigure}[b]{0.4\textwidth}
    \centering
    \[
    \Qcircuit @C=1em @R=1em {
      & \lstick{\ket{0}_{a}}          & \gate{H}      & \ctrl{1}      & \ctrlo{1}\qw & \gate{H}  & \meter & \cw \\
      & \lstick{\ket{0}^{\otimes n}}   & \qw           & \gate{U_1}    & \gate{U_2}   & \qw       & \qw    & \qw
    }
    \]
    \caption{}
    \label{fig:sswapt}
  \end{subfigure}
  \hspace{1cm}
  \begin{subfigure}[b]{0.4\textwidth}
    \centering
    \[
    \Qcircuit @C=1em @R=1em {
      & \lstick{\ket{0}_{a}}          & \gate{H}      & \qw           & \ctrl{1}            & \gate{H}  & \meter & \cw \\
      & \lstick{\ket{0}^{\otimes n}}   & \qw           & \gate{U_1}    & \multigate{1}{SWAP} & \qw       & \qw    & \qw \\
      & \lstick{\ket{0}^{\otimes n}}   & \qw           & \gate{U_2}    & \ghost{SWAP}        & \qw       & \qw    & \qw
    }
    \]
    \caption{}
    \label{fig:swapt}
  \end{subfigure}
  \caption{Variational quantum circuit architectures for knowledge graph embedding. (a) Circuit implementing the SWITCH test to estimate the score function. (b) Circuit implementing the standard SWAP test for score computation.}
  \label{fig:circuitos}
\end{figure}
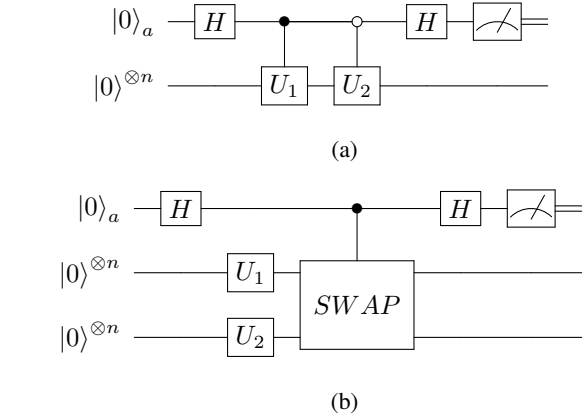

A first approach by Ma et al.~\cite{Ma2019} uses $n+1$ qubits. The head $h$ and tail $t$ entities, along with the relation $r$, are encoded via unitaries $U_1 = U(\vec{\theta}_r)U(\vec{\theta}_h)H^{\otimes n}$ and $U_2 = U(\vec{\theta}_t)H^{\otimes n}$, with the score given by the real part of $\langle t|U(\vec{\theta}_r)| h \rangle$ using a SWITCH test~\cite{Chamorro2023} (see Fig.~\ref{fig:sswapt}):  
\begin{equation}
\delta^{switch}_{hrt} = \Re(\langle t|U(\vec{\theta}_r)| h \rangle) \in [-1,1].
\end{equation}

Another proposal by Kurokawa et al.~\cite{Kurokawa2022} employs $2n+1$ qubits, encoding $U_1$ and $U_2$ in separate $n$-qubit registers, and uses the standard SWAP test~\cite{Buhrman2001} to compute the score (Fig.~\ref{fig:swapt}):  
\begin{equation}
\delta^{swap}_{hrt} = |\langle t|U(\vec{\theta}_r)| h \rangle|^2 \in [0,1].
\end{equation}

Both models use a mean squared error loss,  
\begin{equation}
\mathcal{L} = \min_{\Theta} \frac{1}{|\mathcal{D}|} \sum_{hrt} (\delta_{hrt} - y_{hrt})^2,
\end{equation}
with parameter set $\Theta = (\vec{\theta}_e, \vec{\theta}_r)$.

To achieve quantum advantage, ansatz circuits must remain shallow (with depth scaling linearly in $n$), ensuring that the per-epoch complexity scales as $\mathcal{O}(|\mathcal{D}| \log d / \epsilon^2)$, where $d = 2^n$ is the Hilbert space dimension and $1/\epsilon^2$ is the number of measurements required for score estimation with precision $\epsilon$.

\subsection{A general framework for hybrid KGs embeddings}
Within this context, we have already introduced the general qCUERO framework (Quantum Classical Universal Embeddings for Relations and Objects) \cite{Santesteban2025}, which unifies such approaches by encoding each entity $e \in V$ into a normalized quantum state $|e\rangle = V(\vec{\theta}_e) H^{\otimes n} |0\rangle^{\otimes n}$, where $V(\vec{\theta}_e)$ is a parametrized ansatz over $n$ qubits, and Hadamard gates are applied to initialize the state into a uniform superposition. Relations are represented as unitary operations $U(\vec{\theta}_r)$ acting on the entity states, with trainable parameters $\vec{\theta}_r$.

At the core of this framework lies the assumption that, for a valid triple $(h, r, t)$, the transformed head state should approximate the tail state, that is, $U(\vec{\theta}_r) |h\rangle \approx |t\rangle$. Based on this principle, the hybrid quantum-classical pipeline comprises three main stages: \textit{(i)} quantum state preparation and scoring, \textit{(ii)} classical evaluation of a loss function, and \textit{(iii)} classical optimization of the quantum circuit parameters (see Fig.~\ref{fig:pipeline}). Notably, general results can be established regarding the expressiveness of all models that fit within the qCUERO framework. In particular, the qCUERO model is capable of capturing symmetric and antisymmetric relations, as well as relation inversion and composition~\cite{Santesteban2025}. Among classical embedding models, only RotatE offers a comparable level of expressiveness.

\begin{figure}[!t]
\centering
\includegraphics[width=.55\columnwidth]{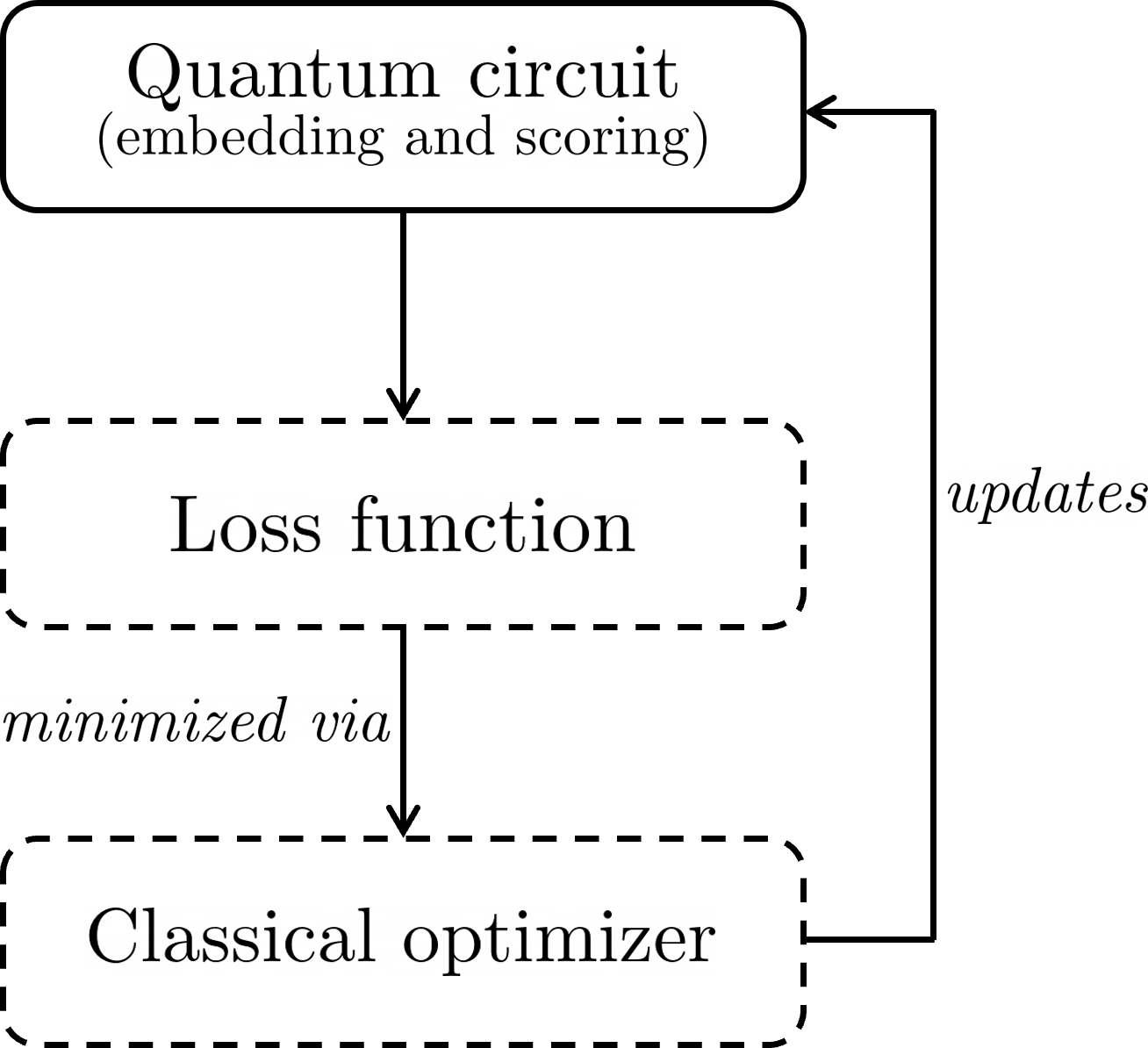}
\caption{General hybrid quantum–classical pipeline for VQA-based knowledge graph embedding. \textit{(i)} Classical inputs are encoded into a parameterized quantum circuit that computes a score; \textit{(ii)} a loss function is evaluated classically using the predicted scores and target values; \textit{(iii)} a classical optimizer updates the parameters of the quantum ansatz. Dashed lines indicate classical computation steps.}
\label{fig:pipeline}
\end{figure}

Both of the VQA-based proposals already discussed~\cite{Ma2019,Kurokawa2022} fit within the general qCUERO framework, differing mainly in the score function used and the number of qubits required. In the next section, we build on this framework to propose a new VQA-based model that avoids the use of ancillary qubits or entangled measurements in score evaluation, improving scalability and implementation feasibility on near-term quantum hardware.

\section{A VQA model for KG embedding without ancillary qubits}
Existing VQA-based approaches for KG embedding exhibit complementary strengths and weaknesses. The proposal by Ma et al.~\cite{Ma2019} employs a single auxiliary qubit and computes a score based on the real part of $\langle t|U(\vec{\theta}_r)| h \rangle$. However, the interpretation of this score is not entirely transparent. For example, it is unclear what a value of $-1$ truly implies in terms of the proximity between entities in the embedding space. On the other hand, the model introduced by Kurokawa et al.~\cite{Kurokawa2022} computes a more natural score for the task, namely the squared overlap $|\langle t|U(\vec{\theta}_r)| h \rangle|^2$, which is easier to interpret and directly aligns with standard similarity measures. Nevertheless, this comes at the cost of requiring $n+1$ auxiliary qubits in order to implement a swap test between $n$-qubit states.

\begin{figure}[!t]
  \centering
    \[
    \Qcircuit @C=1em @R=1em {
      & \lstick{\ket{0}^{\otimes n}}   & \qw           & \gate{U_1}    & \gate{U_2^\dagger}   & \qw   & \meter & \cw
    }
    \]
  \caption{Quantum circuit architecture for the proposed VQA model, using a compute–uncompute scheme to evaluate the score without ancillary qubits.}
  \label{fig:new}
\end{figure}
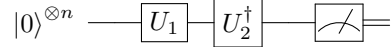

We propose an alternative model within the qCUERO framework that avoids the use of ancillary qubits for computing the score function. Our approach leverages the \emph{compute--uncompute} strategy, which is particularly suitable when the unitaries generating the quantum states of interest are explicitly known (see, e.g.~\cite{havlivcek2019}). The circuit applies $U_1$ followed by $U_2^\dagger$ to the $\ket{0}^{\otimes n}$ state, and the score is estimated from the probability of obtaining the all-zero outcome in a computational basis measurement (Fig.~\ref{fig:new}). This probability corresponds to the squared overlap between the evolved head and tail embeddings,
\begin{equation}
\delta^{c-u}_{hrt} = |\langle t|U(\vec{\theta}_r)| h \rangle|^2 \in [0,1],
\end{equation}
which is the same as $\delta^{swap}_{hrt}$, but without the need for ancillary qubits or entangled measurements.

In comparing the two quantum-enhanced scoring procedures for knowledge-graph embeddings---our compute--uncompute VQA and Kurokawa’s swap-test approach---several practical and statistical trade-offs emerge. On one hand, the compute--uncompute circuit requires only the $n$ system qubits and the inverse unitary $U_2^\dagger$, yielding a shallower depth and eliminating the need for complex controlled-SWAP gates; it directly measures the probability of the all-zero state and thus achieves an $\mathcal{O}(1/\epsilon^2)$ sample complexity without introducing ancilla overhead. However, it relies on the ability to implement $U_2^\dagger$ exactly and suffers from compounded readout errors across all $n$ qubits, since the fidelity of observing $\ket{0}^{\otimes n}$ scales as $F_{\text{read}}^n$, where $F_{\text{read}}$ denotes the single-qubit readout fidelity. By contrast, Kurokawa’s swap test uses a single ancillary qubit---incurring only constant-size readout noise---but at the cost of doubling the register size plus ancilla and implementing multi-qubit controlled-SWAP operations, which increases circuit depth and gate counts. Thus, compute--uncompute excels in resource-limited, low-depth settings where inverse unitaries are accessible, whereas the swap test remains more robust to readout errors at the expense of additional qubits and gate complexity.

In the NISQ era, minimizing the number of controlled gates—particularly multi-qubit gates like CNOTs—is essential to reduce circuit depth and mitigate noise accumulation. Controlled operations are typically more error-prone and contribute disproportionately to decoherence and gate infidelity. Recent studies have emphasized the advantage of circuit designs that limit the use of controlled gates, showing improved performance on current hardware \cite{Wu2020,Wilson2021}.

\section{Concluding remarks}



In this work, we examined the \textit{qCUERO} framework—a variational quantum approach to knowledge graph embedding—and introduced a compute–uncompute-based model that integrates naturally into this setting. The proposed design can represent symmetric, antisymmetric, inverse, and compositional relations. We also showed that earlier methods arise as specific instances within this broader framework.

A key contribution is the formalization of a scoring mechanism built on the compute–uncompute principle, offering a viable alternative to swap-test-based schemes in terms of circuit depth and compatibility with available hardware. A full comparative study of the model's performance against existing approaches is left for subsequent work.

Moving forward, it will be important to test the method on real-world knowledge graphs and to investigate alternative quantum circuit constructions that could further improve embedding efficiency. Another avenue worth exploring is quantum training in superposition, which may help lower the cost of optimization. This study has centered on theoretical aspects. As a next step, we aim to run numerical simulations, for instance to examine how the proposed architecture is affected by noise, and thus gain a clearer view of its performance in practice. Gaining deeper insight into the role of quantum noise will also be essential for developing effective error mitigation strategies tailored to near-term devices.


\printbibliography

@article{Buhrman2001,
  title = {Quantum Fingerprinting},
  author = {Buhrman, Harry and Cleve, Richard and Watrous, John and de Wolf, Ronald},
  journal = {Phys. Rev. Lett.},
  volume = {87},
  issue = {16},
  pages = {167902},
  numpages = {4},
  year = {2001},
  publisher = {American Physical Society},
  doi = {},
}

@article{Chamorro2023,
doi = {},
url = {},
year = {2023},
publisher = {IOP Publishing},
volume = {56},
number = {35},
pages = {355301},
author = {Chamorro-Posada, P and Garcia-Escartin, J C},
title = {The SWITCH test for discriminating quantum evolutions},
journal = {Journal of Physics A: Mathematical and Theoretical}
}

@inproceedings{
sun2018rotate,
title={RotatE: Knowledge Graph Embedding by Relational Rotation in Complex Space},
author={Zhiqing Sun and Zhi-Hong Deng and Jian-Yun Nie and Jian Tang},
booktitle={Int. Conf. on Learning Representations},
year={2019},
url={},
}

@article{Ma2019,
author = {Ma, Yunpu and Tresp, Volker and Zhao, Liming and Wang, Yuyi},
title = {Variational Quantum Circuit Model for Knowledge Graph Embedding},
journal = {Adv. Quantum Tech.},
volume = {2},
number = {7-8},
pages = {1800078},
doi = {},
year = {2019}
}

@INPROCEEDINGS{Kurokawa2022,
  author={Kurokawa, Mori and Giri, Pulak Ranjan and Saito, Kazuhiro},
  booktitle={2022 IEEE Int. Conf. on Quantum Computing and Engineering (QCE)}, 
  title={Evaluating Variational Quantum Circuit Designs for Knowledge Graph Completion}, 
  year={2022},
  volume={},
  number={},
  pages={777-778},
  doi={}}

@article{Hogan2022,
author = {Hogan, Aidan and Blomqvist, Eva and Cochez, Michael and D’amato, Claudia and Melo, Gerard De and others},
title = {Knowledge Graphs},
year = {2021},
issue_date = {May 2022},
publisher = {Association for Computing Machinery},
address = {New York, NY, USA},
volume = {54},
number = {4},
issn = {0360-0300},
url = {},
doi = {},
journal = {ACM Comput. Surv.},
month = jul,
articleno = {71},
numpages = {37}
}

@inproceedings{fan2012graph,
  title={Graph pattern matching revised for social network analysis},
  author={Fan, Wenfei},
  booktitle={Proceedings of the 15th Int. Conf. on Database Theory},
  pages={8--21},
  year={2012}
}

@inproceedings{anand2010techniques,
  title={Techniques for efficiently querying scientific workflow provenance graphs.},
  author={Anand, Manish Kumar and Bowers, Shawn and Lud{\"a}scher, Bertram},
  booktitle={EDBT},
  volume={10},
  number={2010},
  pages={287--298},
  year={2010}
}

@inproceedings{arenas2011querying,
  title={Querying semantic web data with SPARQL},
  author={Arenas, Marcelo and P{\'e}rez, Jorge},
  booktitle={Proceedings of the 30th ACM SIGMOD-SIGACT-SIGART Symp. on Principles of Database Systems},
  pages={305--316},
  year={2011}
}

@article{rossi2021knowledge,
  title={Knowledge graph embedding for link prediction: A comparative analysis},
  author={Rossi, Andrea and Barbosa, Denilson and Firmani, Donatella and Matinata, Antonio and Merialdo, Paolo},
  journal={ACM Trans. on Knowledge Discovery from Data (TKDD)},
  volume={15},
  number={2},
  pages={1--49},
  year={2021},
  publisher={ACM New York, NY, USA}
}

@inproceedings{saeedi2018using,
  title={Using link features for entity clustering in knowledge graphs},
  author={Saeedi, Alieh and Peukert, Eric and Rahm, Erhard},
  booktitle={European Semantic Web Conf.},
  pages={576--592},
  year={2018},
  organization={Springer}
}

@article{kumar2020link,
  title={Link prediction techniques, applications, and performance: A survey},
  author={Kumar, Ajay and Singh, Shashank Sheshar and Singh, Kuldeep and Biswas, Bhaskar},
  journal={Physica A: Statistical Mechanics and its Applications},
  volume={553},
  pages={124289},
  year={2020},
  publisher={Elsevier}
}

@article{ge2024knowledge,
  title={Knowledge graph embedding: An overview},
  author={Ge, Xiou and Wang, Yun Cheng and Wang, Bin and Kuo, C-C Jay and others},
  journal={APSIPA Trans. on Signal and Inf. Proc.},
  volume={13},
  number={1},
  year={2024},
  publisher={Now Publishers, Inc.}
}

@article{dai2020survey,
  title={A survey on knowledge graph embedding: Approaches, applications and benchmarks},
  author={Dai, Yuanfei and Wang, Shiping and Xiong, Neal N and Guo, Wenzhong},
  journal={Electronics},
  volume={9},
  number={5},
  pages={750},
  year={2020},
  publisher={MDPI}
}

@inproceedings{nickel2011three,
  title={A three-way model for collective learning on multi-relational data.},
  author={Nickel, Maximilian and Tresp, Volker and Kriegel, Hans-Peter and others},
  booktitle={Icml},
  volume={11},
  number={10.5555},
  pages={3104482--3104584},
  year={2011}
}

@inproceedings{asmara2023review,
  title={A Review of Knowledge Graph Embedding Methods of TransE, TransH and TransR for Missing Links},
  author={Asmara, Salwana Mohamad and Sahabudin, Noor Azida and Ismail, Nor Syahidatul Nadiah and Sabri, Ily Amalina Ahmad},
  booktitle={2023 IEEE 8th ICSECS},
  pages={470--475},
  year={2023},
  organization={IEEE}
}

@inproceedings{
Santesteban2025,
title={},
author={Martin Santesteban and Patricio Bruno and Santiago Cifuentes and Guidoo Bellomo and Gustavo Bosyk},
booktitle={Argentinean Symposium on Quantum Computing, 54JAIIO},
year={2025},
url={},
}

@article{Wu2020,
  title={QGo: Quantum Circuit Optimization via Graph-Based Reinforcement Learning},
  author={Wu, Xun and Chu, Shidi and Ding, Yujie and Wen, Minghao and Chen, Yuwei and Ding, Yi and Xie, Yuan and Tay, Wee Peng},
  journal={arXiv preprint arXiv:2012.09835},
  year={2020},
  url={https://arxiv.org/abs/2012.09835}
}

@article{Wilson2021,
  title={Just-in-Time Quantum Circuit Optimization: Approximate Circuits and Improved Fidelity},
  author={Wilson, Erik and Gheorghiu, Vlad and Mosca, Michele},
  journal={arXiv preprint arXiv:2107.06701},
  year={2021},
  url={https://arxiv.org/abs/2107.06701}
}

@article{havlivcek2019,
  title={Supervised learning with quantum-enhanced feature spaces},
  author={Havl{\'\i}{\v{c}}ek, Vojt{\v{e}}ch and C{\'o}rcoles, Antonio D and Temme, Kristan and Harrow, Aram W and Kandala, Abhinav and Chow, Jerry M and Gambetta, Jay M},
  journal={Nature},
  volume={567},
  number={7747},
  pages={209--212},
  year={2019},
  publisher={Nature Publishing Group UK London}
}

\end{document}